\documentclass[12pt]{article}
\input epsf.tex
\newlength{\vshift}
\newlength{\hshift}

\usepackage{amssymb}
\usepackage{pstricks}

\usepackage{amsmath}
\usepackage{amsthm}
\usepackage{amsopn}
\def\uno{\mbox{1 \kern-.59em {\rm l}}}
\def\beq{\begin{equation}}
\def\eeq{\end{equation}}
\def\bea{\begin{eqnarray}}
\def\eea{\end{eqnarray}}

 \def\nn{\nonumber}

\def\ga{\gamma}
\def\al{\alpha}
\def\si{\sigma}

\begin{document}

 \vspace*{3cm}

\begin{center}

{\bf{\Large Charged lepton electric dipole moment enhancement in the
Lorentz violated extension of the standard model }}

\vskip 4em{{\bf M. Haghighat$^*$}\footnote{mansour@cc.iut.ac.ir},
{\bf I. Motie$^{*\dagger}$}\footnote{iman@mshdiau.ac.ir } and {\bf
Z. Rezaei${^{*\ddag}}$}\footnote{zahra.rezaei@yazd.ac.ir}\vskip 1em
$^*$Department of Physics, Isfahan University of Technology,\\
Isfahan 84156-83111, Iran\\
$^\dagger$Department of Physics, Mashhad Branch, Islamic Azad
University, Iran\\
$^\ddag$Department of Physics, Yazd University, 89195- 741, Yazd,
Iran }
 \end{center}

 \vspace*{1.9cm}

\begin{abstract}
We consider the Lorentz violated extension of the standard model. In
this framework, there are terms that explicitly  violate
CP-symmetry.  We examine the CPT-even $d_{\mu\nu}$-term to find the
electric dipole moment of charged leptons.  We show that the form
factors besides the momentum transfer, depend on a new
Lorentz-scalar, constructing by $d_{\mu\nu}$ and the four momenta of
the lepton, as well.  Such an energy dependence of the electric
dipole form factor leads to an enhancement of the lepton electric
dipole moment at high energy, even at the zero momentum transfer. We
show that at $\frac{|d|p^2}{m^2_l}\sim 1$ the electric dipole moment
of the charged lepton can be as large as $10^{-14} e\,\,cm$.
\end{abstract}

\newpage

\section{Introduction}
As the electron is a fundamental particle, discovering the nonzero
electron electric dipole moment (eEDM), can unambiguously provide an
experimental test on new physics. In fact, electric dipole moment
(EDM) for fundamental particles violates CP symmetry. Although in
the standard model there is no term which explicitly violates the
CP, through the CKM-phase, a tiny EDM can be produced for all
charged leptons. Therefore, to have the eEDM, comparable with the
experimental bounds, one needs a new theory beyond the standard
model. The new sources of CP-violations in such theories might have
the same origin as the SM. For instance, in the SUSY the electron
EDM originates in new CP-violating phases. In contrast, there might
be theories with explicit CP-violating terms. In the framework of
the standard model extension (SME), introduced by D. Colladay and V.
Alan Kostelecky \cite{SME}-\cite{SME1}, there is such terms. The
phenomenological aspects of the SME have been extensively considered
by many authors in terrestrial
\cite{terrestrial}-\cite{terrestrial13} and astrophysical systems
\cite{astro}-\cite{astro10}, for more than a decade.  The bounds on
the LV-parameters are collected in \cite{table}.  Here we examine
the CPT-even $d_{\mu\nu}$-term that violates the CP symmetry to find
the charged lepton EDM. The electric dipole form factor, as well as
the others, depends not only on the momentum transfer, but also on
this new constant tensor that violates the Lorentz symmetry.
Therefore, one can expect new effects at the zero momentum transfer.
In fact, the form factors should depend on the scalars constructed
by $d_{\mu\nu}$ and four momenta of particles. Therefore, even at
the zero momentum transfer, the form factors may depend on the
energy of the particle and some enhancement for the particle's EDM
with the energy can occur.

In Sec. II, the QED part of the SME and subsequently, the
electromagnetic form factors and their impacts on the charged lepton
EDM are introduced. In Sec. III, we explore the one-loop correction
on the lepton-photon vertex, in the extended QED, and consequently,
the lepton EDM, in high and low energy limits, are obtained. In Sec.
IV, some concluding remarks are given. In appendix A some useful
identities is introduced.  The detail calculations of the vertex
correction is given in appendix B.
\section{Electromagnetic form factors}\label{form factor}
In the QED part of the SME the Lagrangian for a free particle is
parameterized as \cite{SME}-\cite{SME1}
\begin{eqnarray}\label{1}
\mathcal{L}=\bar{\psi}(i\Gamma_{\mu}\partial^{\mu}-M)\psi,
\end{eqnarray}
where
\begin{eqnarray}\label{2}
\Gamma_{\mu}&=&\gamma_{\mu}+c_{\nu\mu}\gamma^{\nu}-d_{\nu\mu}\gamma^{\nu}\gamma^{5}
+e_{\mu}+if_{\mu}\gamma^{5}+\frac{1}{2}g_{\lambda\nu\mu}\sigma^{\lambda\nu},\nonumber\\
M&=&m+a_{\mu}\gamma^{\mu}-b_{\mu}\gamma^{\mu}\gamma^{5}
+\frac{1}{2}H_{\mu\nu}\sigma^{\mu\nu}+im_{5}\gamma^{5}.
\end{eqnarray}
As in \cite{Altschul2006}-\cite{Altschul20061} was noted, the
violating Lorentz parameters in $\Gamma_{\mu}$ are appeared in the
Lagrangian along with a momentum factor and therefore, at high
energy limit, are more important than the LV-parameters which is
given in the mass term $M$. Furthermore, in $\Gamma_{\mu}$, at the
lowest order in the Lorentz violating parameters, only $f_{\mu}$ and
$d_{\mu\nu}$ can produce EDM for point particles. In this article,
we are looking for some enhancement, at high energy limit, on the
EDM of the charged leptons. For this purpose, since $f_{\mu}$ is
unphysical \cite{Altschul2006b}, we restrict ourselves to the
parameter $d_{\mu\nu}$. It should be noted that although the
particle Lorentz transformation symmetry is broken, the Lagrangian
(\ref{1}) is fully covariant under the observer Lorentz
transformations \cite{SME}-\cite{SME1}. Therefore, under the
observer Lorentz transformation, $d_{\mu\nu}$ behaves as a new
Lorentz quantity.
 Consequently, the most general form for the
electromagnetic current between Dirac leptons, consistent with the
Lorentz covariance and the Ward identity, can be written as follows
\begin{eqnarray}
<\psi(p)|J^{EM}_\mu|\psi(p')>&=& \bar{u}(p'){\cal{G_\mu}} (q^2)u(p),
\end{eqnarray}
where $q_\mu = p'_\mu-p_\mu$ and
\begin{eqnarray}
{\cal{G_\mu}}(q^2)&=&F_1\Big[\,\gamma_\mu+\gamma_5\ga^\nu
d_{\nu\mu}\Big]+F_2\,\,i\frac{\sigma_{\mu\nu}q^\nu}{2m}
+F_3\Big[(q_\mu-\frac{q^2}{2m}\gamma_\mu)\gamma_5+\frac{q^2}{2m}d_{\nu\mu}\ga^\nu\Big]\nn\\
&+&F_4\,\, \sigma_{\mu\nu}
\frac{q^\nu}{2m}\gamma_5+{\cal{F}}_d,\label{form-factor}
\end{eqnarray}
in which $m$ is the charged lepton mass and $F_i$'s $i=1-4$ are the
usual electric charge,  magnetic dipole, anapole (axial charge) and
electric dipole form factors, respectively.  Meanwhile,
${\cal{F}}_d$ stands for all the new terms in the current that
vanishes at $d=0$. This part contains the new form factors which can
be defined, for a symmetric and traceless $d_{\mu\nu}$, as follows
\begin{eqnarray}
{\cal{F}}_d &=& (iF_5+F_6\gamma_5)[d_{\mu\alpha}\sigma^{\alpha\nu}-
d^{\nu\alpha}\sigma_{\alpha\mu}]\frac{q_\nu}{2m}\nonumber\\
&+&(F_{7}+F_{8}\gamma_5)[q\cdot d\cdot \gamma q_\mu- q^2
d_{\mu\alpha}\gamma^\alpha ] .\label{nform-factor}
\end{eqnarray}
 All the form factors
are Lorentz scalars and depend on the scalars $q^2$, $p\cdot d\cdot
p'$, $p'\cdot d\cdot p$, $p\cdot d\cdot p$ and $p'\cdot d\cdot p'$.
One can easily see that the electric dipole form factor $F_4$ leads
to a nonzero EDM for a charged lepton as
\begin{eqnarray}
d_e=-\frac{F_4|_{q^2=0}}{2m}. \label{edm}
\end{eqnarray}
It should be noted that in the ordinary standard model
$F_3\Big[(q_\mu-\frac{q^2}{2m}\gamma_\mu)\gamma_5\Big] $ shows the
anapole term in the matrix element of a conserved four-current for a
free spin-$\frac{1}{2}$ fermion \cite{anapole}.  Meanwhile, in the
SME the Dirac equation is modified ( see (\ref{em1})) and therefore
the current conservation leads to a new term for the anapole as
given in (\ref{form-factor}).

 In the Lorentz conserving QED only virtual quarks,
in the loops, can violate the CP-symmetry that in turn, leads to a
tiny nonzero value for $F_4$.  However, in the LV counterpart of the
QED not only $F_4$ is nonzero, even at the leading order,
  but also it depends on the new scalars such as $p\cdot d\cdot p'$ that
  can enhance the lepton's EDM at the high energy limit.  It should be
  noted that the other new form factors given in (\ref{nform-factor}) have also some contribution to the
lepton's EDM as well.  For instance, the $F_6$-term can couple to an
external field $A_{\mu}=(\phi,0,0,0)$ as
\begin{equation}
{\cal{G}}_\mu^{(6)} (q^2)A^\mu=i
F_6\gamma_5(d_{\mu\alpha}\sigma^{\alpha\nu}-
d^{\nu\alpha}\sigma_{\alpha\mu})\frac{q_\nu A^\mu}{2m}=\frac{i
F_6}{2m}\gamma_5(d_{00}\sigma^{0i}+d^{ij}\sigma_{0j}+d_{0j}\sigma^{ji})q_iA_0.
\end{equation}
Meanwhile, in the limit $p$ and $p'\ll m$ one has
\begin{eqnarray}
u_(p)\simeq\sqrt{m}\left(%
\begin{array}{cc}
 (1-\frac{\bf p\cdot\bf\sigma}{2m})\,\,\xi\\
 (1+\frac{\bf p\cdot\bf\sigma}{2m})\,\,\xi\\
\end{array}%
\right),
\end{eqnarray}
therefore, the spin dependent part of the current, at the zero
momentum transfer and up to the first order of the LV-parameter, can
be easily casted into
\begin{equation}
\bar{u}(p'){\cal{G}}_0^{(6)} (q^2)u(p)\simeq
-F_6(0)[d_{00}{\bar\xi}\sigma^i\xi + d_{ji}{\bar\xi}\sigma^j\xi]q_i.
\end{equation}
 It should be noted that since ${\cal{G}}_\mu^{(6)} $ depends on the
LV-parameter, then the spinors in the current, at the first order of
the LV-parameter $d$, are the free ones. In the high energy limit,
the spinors can be given as
\begin{eqnarray}
u_(p)\simeq \frac{\sqrt{2E}}{2}\left(%
\begin{array}{cc}
 (1-\hat{p}\cdot\sigma)\,\,\xi\\
 (1+\hat{p}\cdot\sigma)\,\,\xi\\
\end{array}%
\right),
\end{eqnarray}
though the spin dependent part of the current does not change.
Therefore, the form factor $F_6$ leads to the electric dipole
interaction as
\begin{eqnarray}
-d_e\cdot {\cal{E}}=\frac{eF_6|_{q^2=0}}{m}(d_{00}S\cdot {\cal{E}}
+S_id_{ij}{\cal{E}}_j). \label{edm1}
\end{eqnarray}
Before calculating the form factors, some comments are in order. As
(\ref{nform-factor}) shows, the Lorentz vector ${\cal{F}}_d$ is
constructed by the Lorentz tensor $d_{\mu\nu}$.  Therefore, up to
the first order of $d_{\mu\nu}$, only the form factors $F_1$-$F_4$
depend on the LV-parameter. In fact, at the leading order, all the
new form factors are $d$-independent and, at the zero recoil, they
are of the order of $\frac{\alpha}{2\pi}$.  Thus, the form factors
such as the $F_6$, see Eq.(\ref{edm1}), lead to $d_e\sim
\frac{\alpha}{2\pi} \frac{e|d|}{2m_e}$ or $|d|\sim 10^{-14}$ for the
eEDM of the order of $10^{-27} e\,\,cm$. Meanwhile, at the leading
order, $F_4|_{q^2=0}\sim \frac{\alpha}{2\pi}
\frac{p_id_{ij}p_j}{m^2_e}$ that in turn results in $d_e\sim
\frac{\alpha}{2\pi} \frac{e|d|p^2}{2m^3_e}$.  In the other words, in
the relativistic limit, there is an enhancement on the eEDM through
the form factor $F_4$.
 It should be noted that, in any case it is assumed that $\frac{|d|p^2}{m^2_e}\leq 1$ and the LV parameter $d_{\mu\nu}$
 is symmetric and can be taken traceless \cite{SME}-\cite{SME1}.
\section{Charged Lepton EDM in the
standard model extension }\label{vc} To obtain the $F_i$'s in the
electromagnetic current, we examine the lepton-photon vertex in the
QED part of the SME.  In this section, as a crosscheck,  we assume
both symmetric and antisymmetric parts of $d_{\mu\nu}$ are nonzero,
however, at the end we show that the lepton EDM as a physical
quantity depends only on the symmetric part of $d_{\mu\nu}$.  The
effective lagrangian for the only non vanishing LV-parameter
$d_{\mu\nu}$ is
\begin{eqnarray}
{\cal{L}}^{CPT-even}_{electron} &=&\frac{i}{2} \bar{\psi} \gamma^\mu
\overleftrightarrow{D}_{\mu} \psi -
m\bar{\psi}\psi+\frac{i}{2}\:d_{\mu\nu}\bar{\psi}\gamma_{5}
\gamma^{\mu}\overleftrightarrow{D}^{\nu}\psi. \label{lagrangian}
\end{eqnarray}
 The Lagrangian (\ref{lagrangian}) leads to the equation of
motion for a free lepton as
\begin{eqnarray}
(\not \!p- m+d_{\mu\nu}p^\nu\gamma_5\gamma^\mu)u(p)&=&0. \label{em1}
\end{eqnarray}
Meanwhile, the modified Gordon identities can be obtained as follows
 \bea
\bar{u}\ga_{\mu}u=\bar{u}\frac{(p+p')_{\mu}}{2m}u+\bar{u}\frac{i\si_{\mu\nu}q^{\nu}}{2m}u+
\bar{u}i\frac{\si_{\mu\al}d^{\al\nu}(p+p')_\nu}{2m}\ga_5u+\bar{u}\frac{d_{\mu\nu}q^{\nu}}{2m}\ga_5u,
\eea and \bea
\bar{u}\ga_{\mu}\ga_5u=\bar{u}\frac{q_{\mu}\ga_5}{2m}u+\bar{u}\frac{i\si_{\mu\nu}(p+p')^{\nu}\ga_5}{2m}u+
\bar{u}i\frac{\si_{\mu\al}d^{\al\nu}q_\nu}{2m}u+\bar{u}\frac{d_{\mu\al}(p+p')^{\al}}{2m}u.
\eea  Therefore, to the leading order of the LV-parameter
$d^{\mu\nu}$, the lepton-photon vertex
 $ie(\gamma^\mu+d^{\nu\mu}\gamma_{5}\gamma_{\nu})$ can be written as
\begin{eqnarray}
\bar{u}(\ga_{\mu}+\ga_5\ga^\nu d_{\nu\mu})u
&=&\bar{u}(\frac{(p+p')_{\mu}}{2m}+\frac{i\si_{\mu\nu}q^{\nu}}{2m})u\nn\\
&+& \bar{u}[\frac{(d_{\mu\nu}-d_{\nu\mu})q^{\nu}}{2m}+
i\frac{(\si_{\mu\al}d^{\al\nu}+\si^{\nu\al}d_{\al\mu})(p+p')_\nu}{2m}]
\ga_5u.\label{tree}
\end{eqnarray}
In (\ref{tree}) the antisymmetric tensor
$d_{\mu\nu}^A=(d_{\mu\nu}-d_{\nu\mu})$ can couple to an electric
field as $d^A_{i 0}{\cal{E}}_i$ which is a constant and, as is
expected, it has not any contribution to the EDM.  Meanwhile, to
avoid a nonstandard time derivatives in the canonical quantization
procedure of the fermion fields, $\Gamma_0$ in (\ref{2}) must be
equal to $\gamma_0$ or $d_{\mu 0}=0$
\cite{Altschul2006}-\cite{Altschul20061}.  In fact, to support
$\Gamma_0=\gamma_0$ in (\ref{1}), one needs a field redefinition
$\psi=A\chi$ \cite{redefinition}-\cite{redefinition1} where its
existence was shown in \cite{existence} and is given in \cite{A} for
$\Gamma_0=c_{\nu 0}\gamma^\nu$.  In our case, to leading order of
$d_{\mu\nu}$, we introduce $A=1+\frac{1}{2}d_{\mu
0}\ga^0\ga^\mu\ga_5$. Therefore, the lagrangian (\ref{lagrangian})
in terms of the new field $\chi$ transforms into
\begin{eqnarray}
{\cal{L}}^{CPT-even}_{electron} &=&\frac{i}{2}
\bar{\chi}{\tilde{\eta}}_{\mu\nu} \gamma^\mu
\overleftrightarrow{D}^{\nu} \chi - {\tilde{m}}\bar{\chi}\chi,
\label{lagrangian1}
\end{eqnarray}
where
\begin{eqnarray}
{\tilde{\eta}}_{\mu\nu} &=& {\eta}_{\mu\nu}+ {\cal{D}}_{\mu\nu}\ga_5, \nonumber\\
{\cal{D}}_{\mu\nu}&=&
d_{\mu\nu}-\eta_{\mu\nu}d_{00}+\eta_{0\mu}d_{\nu
0}-\eta_{0\nu}d_{\mu 0},\nonumber\\
\tilde{m} &=& m(1+id_{\al 0}\si^{\al 0}\ga_5).
\end{eqnarray}
One can easily see that ${\cal{D}}_{\mu 0}=0$.  Here, for
simplicity, we assume $d_{\mu 0}=0$ then ${\cal{D}}_{\mu \nu}=d_{\mu
\nu}$, $\tilde{m}= m$ and the fermion propagator for the new field
is \beq S_F(p)=\frac{1}{{\tilde{\eta}}_{\mu\nu}\ga^\mu p^\nu-m}.
\label{propagator}\eeq

Since the electromagnetic current, at the tree level, has not any
contribution on the lepton EDM, then to find a nonzero value for the
EDM we consider the one loop correction on the lepton-photon vertex
in the framework of the QED part of SME. As is shown in Fig.~1,
there are five places which are affected by the LV parameter $d$.
To evaluate the one loop correction in the QED extension (QEDE), one
has
\begin{eqnarray}
\Gamma^{\mu}_{QEDE}=\int
\frac{d^{4}k}{(2\pi)^4}\frac{-ig_{\rho\alpha}}{(p-k)^{2}}\bar{u}(p')(-ie\Gamma^{\alpha})S_F(k')\Gamma^{\mu}S_F(k)(-ie\Gamma^{\rho})u(p),
\label{vertex-function}
\end{eqnarray}
in which $\Gamma^{\mu}_{QEDE}=\bar{u}(p'){\cal{G^\mu}} (q^2)u(p)$,
$S_F$ is given in (\ref{propagator}) and
$\Gamma^{\mu}=(\gamma^\mu+d^{\nu\mu}\gamma_{5}\gamma_{\nu})$.
Replacing $S_F$ with its expansion up to the first order of $d$ cast
the vertex function into
\begin{eqnarray}
\Gamma^{\mu}_{QEDE}&=&\int
\frac{d^{4}k}{(2\pi)^4}\frac{-ie^2}{(p-k)^{2}}\bar{u}(p')\{\Gamma^{\alpha}\frac{(\!\not{k'}+m)}{k'^2-m^2}\Gamma^{\mu}
\frac{(\!\not{k}+m)}{k^2-m^2}\Gamma_{\alpha}\nonumber\\
&+&\Gamma^{\alpha}\frac{(\!\not{k'}+m)}{k'^2-m^2}\Gamma^{\mu}
\frac{(\!\not{k}+m)}{k^2-m^2}\gamma\cdot d\cdot k
\gamma_5\frac{(\!\not{k}+m)}{k^2-m^2}\Gamma_{\alpha}\nonumber\\
&+&\Gamma^{\alpha}\frac{(\!\not{k'}+m)}{k'^2-m^2}\gamma\cdot d\cdot
k' \gamma_5\frac{(\!\not{k'}+m)}{k'^2-m^2}\Gamma^{\mu}
\frac{(\!\not{k}+m)}{k^2-m^2}\Gamma_{\alpha}\}u(p).
\label{vertex-function}
\end{eqnarray}

\begin{figure}[t]
\centerline{\epsfysize=4in\epsfxsize=4in\epsffile{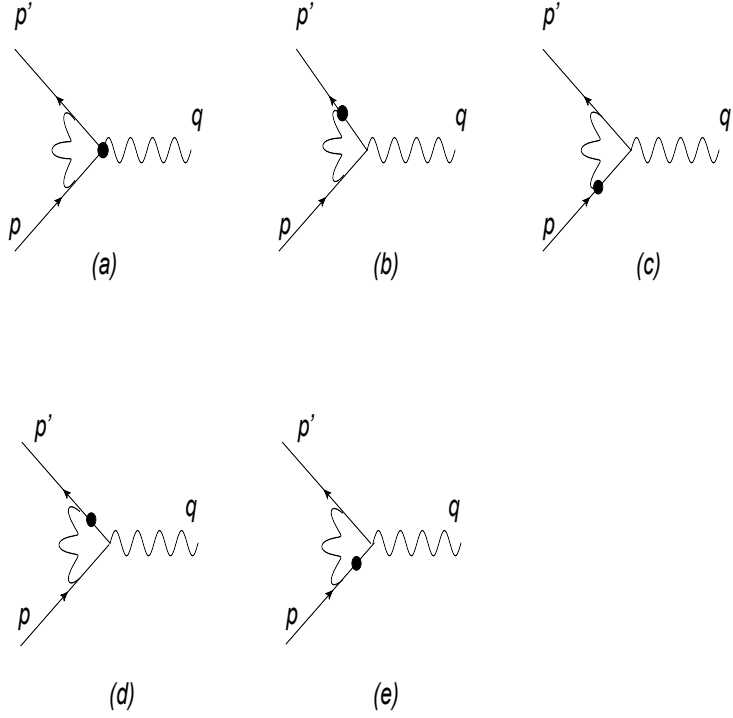}}
\caption{\scriptsize\tiny
 The one loop diagrams for lepton-photon vertex in the extended QED up to the first order of the lorentz violation
parameter. The solid circle on each diagram shows the first order
LV-contribution from the extended QED. a-c represent the
LV-correction on the vertex while d and e show the corrected
propagators. } \label{vexf}
\end{figure}
As is already mentioned, the most important term for the EDM, in the
high energy limit, is $F_4$.  In fact, this form factor at the zero
recoil depends on the scalar $p\cdot d\cdot p$ which enhances the
value of the EDM in the higher energies. Therefore, to evaluating
the vertex function, we only retain those terms which are
proportional to $p\cdot d\cdot p$.  To simplify
(\ref{vertex-function}), we introduce two identities as follows
\begin{eqnarray}
\Gamma^{\alpha}d_{\mu\nu}\gamma^\nu\gamma_5\Gamma_{\alpha}&=&\gamma^{\alpha}d_{\mu\nu}\gamma^\nu\gamma_5\gamma_{\alpha}\nonumber\\
&=&2d_{\mu\nu}\gamma^\nu\gamma_5, \label{identity1}
\end{eqnarray}
and
\begin{eqnarray}
\Gamma^{\alpha}\Gamma_\mu\Gamma_{\alpha}&=&\Gamma^{\alpha}\gamma_\mu\Gamma_{\alpha}-\Gamma^{\alpha}d_{\mu\nu}\gamma^\nu\gamma_5\Gamma_{\alpha}\nonumber\\
&=&2
\gamma_\alpha(d_{\mu\alpha}+d_{\alpha\mu})-2d_{\mu\nu}\gamma^\nu\gamma_5-2\gamma_\mu(1+d^\alpha_\alpha).
\label{identity2}
\end{eqnarray}
Then, the expression for $F_4$, after manipulating some algebra that
is given in appendix B, can be obtained at the zero recoil and up to
the first order of $d$, as in
\begin{eqnarray}
F_4&=&\frac{275\al}{18\pi}\Bigg\{\frac{p.d^S.p}{m^2} \Bigg\},
\label{f44}
\end{eqnarray}
in which $d^S$ is the symmetric part of the LV parameter
$d_{\mu\nu}$. Consequently, one finds
\begin{eqnarray}
d_e&=&7\times10^{-13}\frac{p\cdot d^S\cdot p}{m_e^2}\,\,\, e\,\,cm\,
, \label{eEDM}
\end{eqnarray}
 for the electron's EDM and
\begin{eqnarray}
d_\mu&=&3\times10^{-15}\frac{p\cdot d^S\cdot p}{m_\mu^2}\,\,\,
e\,\,cm\, , \label{muEDM}
\end{eqnarray}
for the muon's EDM.  One should note that, at the low energy limit
where the EDM of the electron as a stable particle is measured, the
correction given in (\ref{eEDM}) in comparison with (\ref{edm1}) is
irrelevant.  In contrast, the heavy charged leptons due to their
short lifetimes should be measured in apparatus like the storage
ring, as is suggested in \cite{storage} for the charged leptons and
in \cite{storage1} for the other heavy charged particles. Therefore,
for instance, (\ref{muEDM})  can be used to put an upper bound on
the LV-parameter $d$ for the muon.  In the storage ring, muons are
in the xy plane therefore, besides $p_z=0$ both $p_x$ and $p_y$, in
average, are equal to zero.  Therefore, at the high energy limit
(\ref{muEDM}) leads to

\begin{eqnarray}
d_\mu&=&3\times10^{-15}\frac{p_0^2(d_{xx}+d_{yy})}{2m_\mu^2}\,\,\,
e\,\,cm\, . \label{muEDM1}
\end{eqnarray}
To compare  (\ref{muEDM1}) with different experiments, it is
convenient to use the standard Sun-centered inertial reference frame
\cite{sun}-\cite{sun1}.  Denote a non rotating basis by $( X; Y;
Z)$, with $Z$ parallel to the earth's axis along the north direction
and the $X$ and $Y$ axes lying in the plane of the earth's equator.
Thus, the quantity $d_{xx}+d_{yy}$ in this frame is
\begin{eqnarray}
d_{xx}+d_{yy}&=&(1-\sin^2\chi\cos^2\Omega
t)d_{XX}-\frac{1}{2}\sin^2\chi\sin2\Omega t(d_{XY}+d_{YX})\nonumber\\
&-&\frac{1}{2}\sin2\chi\cos\Omega
t(d_{XZ}+d_{ZX})-\frac{1}{2}\sin2\chi\sin\Omega
t(d_{YZ}+d_{ZY})\nonumber\\&+&(1-\sin^2\chi\sin^2\Omega
t)d_{YY}+\sin^2\chi d_{ZZ}, \label{rt}
\end{eqnarray}
where $\chi$ is the geographic colatitude of the experiment
location.  As (\ref{rt}) shows the $\mu EDM$ is a time dependent
quantity.  Meanwhile, the time average of (\ref{rt}) leads to
\begin{eqnarray}
d_{xx}+d_{yy}&=&(d_{XX}+ d_{YY}) -\frac{1}{2}\sin^2\chi (d_{XX}+
d_{YY}-2 d_{ZZ}), \label{rti}
\end{eqnarray}
where for measurements made at different $\chi$ one has
\begin{eqnarray}
\delta(d_{xx}+d_{yy})&=&\frac{1}{2}(\sin^2\chi_1-\sin^2\chi_2
)(d_{XX}+ d_{YY}-2 d_{ZZ}). \label{rtid}
\end{eqnarray}

 The experimental bound on the $\mu$EDM is about $1.8\times 10^{-19}
\,e\,cm$  \cite{muEDM} where $\chi=49.1$ for the E821 experiment and
the muon energy is of the order of $3\,GeV$.  Therefore,
(\ref{muEDM1}) and (\ref{rti}) results in
\begin{eqnarray}
d_\mu&=&1.2\times 10^{-12}[0.71(d_{XX}+ d_{YY})
 +0.57d_{ZZ}]\,\,\,
e\,\,cm\, , \label{muEDM2}
\end{eqnarray}
or
\begin{eqnarray}
[0.71(d_{XX}+ d_{YY}) + 0.57d_{ZZ}]<1.5\times 10^{-7}, \label{bound}
\end{eqnarray}
which is the first bound on the combination of $d_{ii}$ components
of the Lorentz violation parameter $d$ for muon.  One should note
that to see the enhancement on the eEDM at the high energy limit one
needs to examine an indirect experiment such as $e^-e^+\rightarrow
l^-l^+$ at the LEP. As was shown in \cite{ee-ll}, the EDM of leptons
about $10^{-17}\,e\,cm \sim 10^{-3}\,GeV^{-1}$ may have some
measurable contribution on the $e^-e^+\rightarrow l^-l^+$ which is
comparable to the interference term coming from the one $Z$-boson
exchange channel. In fact, besides the ordinary one photon exchange
diagram, there are diagrams at the lowest order in which one of the
vertices is replaced by the electric-dipole one. Therefore, for the
non vanishing interference term, there is an extra power of the
momentum in the amplitude and the fractional correction with respect
to the ordinary QED is of the order of $d_l E$ where $d_l$ is the
$l$EDM. This correction is about 20 percent for $d_l\sim
10^{-3}\,GeV^{-1}$ and $E\sim 200 \,GeV$ . Unfortunately, the
interference term is zero and the fractional correction is of the
order of $(d_l E)^2\sim .02$. Consequently, the LV-parameter
$d=8.9\times10^{-17}$ for the electron leads to a few percent
fractional correction to the $e^-e^+\rightarrow l^-l^+$. Meanwhile,
since $m_\mu\sim 200 m_e$ then to have the same order of magnitude
correction, through the $\mu$EDM, the LV-parameter $d$ for the muon
should be $ 9\times 10^{-10}$.


\section{Conclusion }
We examined the electric dipole moment of the charged fermions in
the QED part of the SME.  Besides the ordinary form factors there
are a lot of new form factors in the SME framework, see
(\ref{form-factor}). In addition to the $q^2$, the ordinary form
factors, up to the first order of the LV parameter, depend on new
Lorentz scalars such as $p.d.p$, see (\ref{f44}). Meanwhile, the new
form factors, to the leading order of the LV-parameter $d$, depend
only on the $q^2$, see (\ref{edm1}).
 Therefore, the ordinary form factors in contrast with the QED
counterpart, at the zero momentum transfer, depend on the energy of
the particles, see (\ref{f44}). The energy dependence of the form
factors lead to an enhancement of the electric dipole moment of
leptons at high energy limit, see (\ref{eEDM}). In fact, at the high
energy limit, but low enough to satisfy $\frac{|d|p^2}{m^2_e}\leq
1$, the eEDM can be as large as $\sim 10^{-14}\,\, e\,cm$, see
(\ref{eEDM}). Consequently, the LEP data can be used to put bounds
on $d$, via the enhanced EDM, of the order of $9\times10^{-17}$ and
$9\times10^{-10}$ for the electron and muon, respectively. Using the
storage ring data for the muon, a bound on $[0.71(d_{XX}+ d_{YY}) +
0.57d_{ZZ}]\sim1.5\times 10^{-7}$ has been obtained for the
mu-lepton. In fact, this is the first bound on the components
$|d_{ij}|$ of the muon \cite{table}.

\section{Appendix A }
Here we introduce some useful identities.  The Dirac equation in the
SME is
\begin{eqnarray}
(\not \!p- m+d_{\mu\nu}p^\nu\gamma_5\gamma^\mu)u(p)&=&0,
\end{eqnarray} and
\begin{eqnarray}
\bar{u}(p)(\not \!p- m+d_{\mu\nu}p^\nu\gamma_5\gamma^\mu)&=&0.
\end{eqnarray}
 These equations can be easily casted into
\begin{eqnarray}
\bar{u}(p')(\not \!q) u(p)&=&-\bar{u}(p')(\ga_5\ga.d.q) u(p),
\end{eqnarray} and
\begin{eqnarray}
\bar{u}(p')(\not \!q\ga_5) u(p)&=&\bar{u}(p')(2m\ga_5+\ga.d.q) u(p).
\end{eqnarray}
Also one has
\begin{eqnarray}
p^2u(p)&=&( m^2-2md_{\mu\nu}p^\mu\gamma_5\gamma^\nu+
2p.d.p\gamma_5)u(p),
\end{eqnarray} and
\begin{eqnarray}
\bar{u}(p)p^2&=&\bar{u}(p)( m^2-2md_{\mu\nu}p^\mu\gamma_5\gamma^\nu-
2p.d.p\gamma_5).
\end{eqnarray}
 The Gordon identity for a Dirac particle in a
LV-background $d_{\mu\nu}$ can be obtained as
 \bea
\bar{u}\ga_{\mu}u=\bar{u}\frac{(p+p')_{\mu}}{2m}u+\bar{u}\frac{i\si_{\mu\nu}q^{\nu}}{2m}u+
\bar{u}i\frac{\si_{\mu\al}d^{\al\nu}(p+p')_\nu}{2m}\ga_5u+\bar{u}\frac{d_{\mu\nu}q^{\nu}}{2m}\ga_5u,
\eea and
 \bea
\bar{u}\ga_{\mu}\ga_5u=\bar{u}\frac{q_{\mu}\ga_5}{2m}u+\bar{u}\frac{i\si_{\mu\nu}(p+p')^{\nu}\ga_5}{2m}u+
\bar{u}i\frac{\si_{\mu\al}d^{\al\nu}q_\nu}{2m}u+\bar{u}\frac{d_{\mu\nu}(p+p')^{\nu}}{2m}u.
\eea Some other useful identities are \bea
\bar{u}[\sigma^{\mu\nu}q_\nu
\ga_5]u=i\bar{u}(p+p')^{\mu}\ga_5u+i\bar{u}d^{\mu\nu}q_{\nu}u-
\bar{u}\si^{\mu\al}d_{\al\nu}(p+p')^\nu u,\label{edm} \eea and \bea
\bar{u}[\sigma^{\mu\nu}(p+p')_\nu]
u=i\bar{u}q^{\mu}u+i\bar{u}d^{\mu\nu}(p+p')_{\nu}\ga_5u-
\bar{u}\ga_5\si^{\mu\al}d_{\al\nu}q^\nu u. \eea

\section{Appendix B }
In this appendix we give the details of the vertex function
calculations.  As a crosscheck,  we assume both symmetric and
antisymmetric parts of $d_{\mu\nu}$ are nonzero, however, at the end
we show that the lepton EDM as a physical quantity depends only on
the symmetric part of $d_{\mu\nu}$.  To this end, the equation
(\ref{vertex-function}) can be written as follows
\begin{equation}
\Gamma^{\mu}_{QEDE}=\Gamma_1+\Gamma_2+\Gamma_3,
\end{equation} where
\begin{equation}
\Gamma_1=\int
\frac{d^{4}k}{(2\pi)^4}\frac{-ie^2}{(p-k)^{2}}\bar{u}(p')\{\Gamma^{\alpha}\frac{(\!\not{k'}+m)}{k'^2-m^2}\Gamma^{\mu}
\frac{(\!\not{k}+m)}{k^2-m^2}\Gamma_{\alpha}\}u(p), \label{vf1}
\end{equation}
\begin{equation}
\Gamma_2=\int
\frac{d^{4}k}{(2\pi)^4}\frac{-ie^2}{(p-k)^{2}}\bar{u}(p')\{
\Gamma^{\alpha}\frac{(\!\not{k'}+m)}{k'^2-m^2}\Gamma^{\mu}
\frac{(\!\not{k}+m)}{k^2-m^2}\gamma\cdot d\cdot k
\gamma_5\frac{(\!\not{k}+m)}{k^2-m^2}\Gamma_{\alpha}\}u(p),
\label{vf2}
\end{equation} and
\begin{equation}
\Gamma_3=\int
\frac{d^{4}k}{(2\pi)^4}\frac{-ie^2}{(p-k)^{2}}\bar{u}(p')\{\Gamma^{\alpha}\frac{(\!\not{k'}+m)}{k'^2-m^2}\gamma\cdot
d\cdot k' \gamma_5\frac{(\!\not{k'}+m)}{k'^2-m^2}\Gamma^{\mu}
\frac{(\!\not{k}+m)}{k^2-m^2}\Gamma_{\alpha}\}u(p).
\label{vf3}
\end{equation}
Now, we use the identities (\ref{identity1}) and (\ref{identity2})
to simplify  $ \Gamma_i$'s as follows
\begin{eqnarray}
\Gamma_1&=&\int
\frac{d^{4}k}{(2\pi)^4}\frac{2ie^2}{(p-k)^{2}(k'^2-m^2)(k^2-m^2)}\nonumber\\&
&\bar{u}(p')\{\!\not{k}\gamma_\mu\!\not{k'}
-d^\alpha_\alpha\!\not{k'}\gamma_\mu\!\not{k}\gamma_5-d_{\mu\alpha}\!\not{k}\gamma_\mu\!\not{k'}\gamma_5-
d^s_{\alpha\beta}k^\beta\!\not{k'}\gamma_\mu\gamma^\alpha\gamma_5
\nonumber\\
&+& d^s_{\alpha\mu}\!\not{k'}\!\not{k}\gamma^\alpha\gamma_5-
d^s_{\alpha\beta}k'^\beta\gamma_\mu\!\not{k}\gamma^\alpha\gamma_5-2m(k'+k)_\mu-2m
d_{\mu\beta}q^\beta\gamma_5\nonumber\\
&-&2m
i\sigma_{\alpha\beta}d^{\alpha\beta}q_\mu\gamma_5-m\!\not{q}d^A_{\mu\alpha}\gamma^\alpha\gamma_5+
mq^\beta
d^A_{\beta\alpha}\gamma_\mu\gamma^\alpha\gamma_5-m^2d^\alpha_\alpha\gamma_\mu\gamma_5
\nonumber\\
&+&m^2(\gamma_\mu+d_{\mu\alpha}\gamma^\alpha\gamma_5)+m^2d^s_{\alpha\mu}\gamma^\alpha\gamma_5\}u(p),
\label{vf12}
\end{eqnarray}
\begin{eqnarray}
\Gamma_2&=&\int
\frac{d^{4}k}{(2\pi)^4}\frac{-2ie^2}{(p-k)^{2}(k'^2-m^2)(k^2-m^2)^2}\nonumber\\&
&\bar{u}(p')\{\gamma\cdot d\cdot
k\gamma^\mu\!\not{k'}(k^2+m^2)-2\!\not{k}\gamma^\mu\!\not{k'}k\cdot
d^s\cdot k-2mk^2d_{\mu\alpha}k^\alpha\nonumber\\&-& 2m(\gamma\cdot
d\cdot
k\!\not{k'}\gamma^\mu\!\not{k}+\!\not{k}\gamma^\mu\!\not{k'}\gamma\cdot
d\cdot k)+4m(k'+k)_\mu k\cdot d^s\cdot k\nonumber\\
&+& 2m^2(\gamma\cdot d\cdot k\!\not{k}\gamma^\mu-\gamma_\mu k\cdot
d^s\cdot k)-2m^3d_{\mu\alpha}k^\alpha\}\gamma_5u(p), \label{vf22}
\end{eqnarray}
and
\begin{eqnarray}
\Gamma_3&=&\int
\frac{d^{4}k}{(2\pi)^4}\frac{-2ie^2}{(p-k)^{2}(k'^2-m^2)^2(k^2-m^2)}\nonumber\\&
&\bar{u}(p')\{\!\not{k}\gamma^\mu \gamma\cdot d\cdot
k'(k'^2+m^2)-2\!\not{k}\gamma^\mu\!\not{k'}k'\cdot d^s\cdot
k'\nonumber\\&+&2m(k'^2+m^2)d_{\mu\alpha}k'^\alpha +2m(\!\not{k}
\gamma\cdot d\cdot
k'\!\not{k'}\gamma^\mu+\gamma^\mu\!\not{k'}\gamma\cdot d\cdot
k'\!\not{k})\nonumber\\&-& 4m k'\cdot d^s\cdot k' q_\mu -8m k'\cdot
d^s\cdot k' k_\mu -2m^2 k'\cdot d^s\cdot k'
\gamma^\mu\nonumber\\&+&2m^2 \gamma^\mu\!\not{k'}\gamma\cdot d\cdot
k'\}\gamma_5u(p), \label{vf32}
\end{eqnarray}
Now, we evaluate the integrals using the standard procedures. We use
the method of Feynman parameters to rewrite the denominators as
follows
\begin{eqnarray}
\frac{1}{(p-k)^{2}(k'^2-m^2)(k^2-m^2)^2}=\int
dxdydz\delta(x+y+z-1)\frac{6x}{D^4},
\end{eqnarray}
and
\begin{eqnarray}
\frac{1}{(p-k)^{2}(k'^2-m^2)^2(k^2-m^2)}=\int
dxdydz\delta(x+y+z-1)\frac{6y}{D^4},
\end{eqnarray}
where $D=l^2-\Delta+i\epsilon$ and
\begin{eqnarray}
\Delta=(1-z)^2m^2-xyq^2 \quad,\quad l=k-zp+yq. \label{delta}
\end{eqnarray}
Here we are interested in the momentum dependent part of the $F_4$
form factor.  Meanwhile, Eq.(\ref{edm}) shows that the  $F_4$ comes
as the coefficient of $(p+p')^{\mu}\ga_5$.  Therefore, we only
retain those momentum dependent terms, in $\Gamma_1$ to $\Gamma_3$,
which are proportional to $(p+p')^{\mu}\ga_5$.  One can see that
only $\Gamma_2$ and $\Gamma_3$ have such terms which after
performing the integrals on the momenta they can be obtained as
follows

\begin{eqnarray}
\Gamma_{2\,p\cdot d\cdot p}&=&\frac{2e^2}{(4\pi^2)}\int
dxdydz\delta(x+y+z-1)\frac{x}{\Delta^2}\bar{u}(p')\{-2[-my(1-y)\nonumber\\&+&
m(z+y)(z-2y+2)][y^2q\cdot d^s\cdot q-2zyq\cdot d^s\cdot p+z^2 p\cdot
d^s\cdot p]\nonumber\\&-& 2m[2y^2zp'\cdot d^s\cdot
p'+2(z+y)^2zp\cdot d^s\cdot p-4yz(z+y)p'\cdot d^s\cdot
p]\nonumber\\&+& 4m[y^2q\cdot d^s\cdot q-2yzq\cdot d^s\cdot
p+z^2p\cdot d^s\cdot p]z \}(p+p')^{\mu}\ga_5u(p),\label{vf23}
\end{eqnarray}
where at $q^2=0$ is
\begin{eqnarray}
\Gamma_{2\,p\cdot d\cdot p}&=&\frac{2e^2}{(4\pi^2)}\int
dxdydz\delta(x+y+z-1)\frac{x}{\Delta^2(q^2=0)}\nonumber\\&
&\bar{u}(p')\{-2m[-y(1-y)+ (z+y)(z-2y+2)]z^2 \nonumber\\&-&
2m[2y^2z+2(z+y)^2z-4yz(z+y)]\nonumber\\&+& 4mz^3 \}p\cdot d^s\cdot
p(p+p')^{\mu}\ga_5u(p),\label{vf230}
\end{eqnarray}
 and
\begin{eqnarray}
\Gamma_{3\,p\cdot d\cdot p}&=&\frac{2e^2}{(4\pi^2)}\int
dxdydz\delta(x+y+z-1)\frac{y}{\Delta^2}\bar{u}(p')\{\nonumber\\&-
&2m[y(z-y+1)+ z(z-2y+2)][z^2p\cdot d^s\cdot p+(1-y)^2q\cdot d^s\cdot
q\nonumber\\&+&2z(1-y)q\cdot d^s\cdot q]+2m[2(z+y)(1-y)^2p'\cdot
d^s\cdot p'\nonumber\\&+& 2(z+y-1)^2(1-y)p\cdot d^s\cdot
p+4z(1-y)(z+y-1)p'\cdot d^s\cdot p]
\nonumber\\&-&4mz((1-y)q+zp)\cdot d^s\cdot((1-y)q+zp)
\}(p+p')^{\mu}\ga_5u(p),\label{vf33}
\end{eqnarray}
where at $q^2=0$ one has
\begin{eqnarray}
\Gamma_{3\,p\cdot d\cdot p}&=&\frac{2e^2}{(4\pi^2)}\int
dxdydz\delta(x+y+z-1)\frac{y}{\Delta^2(q^2=0)}\bar{u}(p')\{-2m[y(z-y+1)\nonumber\\&+&
z(z-2y+2)]z^2+2m[2(z+y)(1-y)^2+
2(z+y-1)^2(1-y)\nonumber\\&+&4z(1-y)(z+y-1)]-4mz^3\}p\cdot d^s\cdot
p(p+p')^{\mu}\ga_5u(p),\label{vf330}
\end{eqnarray}
where the subscript $p\cdot d\cdot p$ stands for the momentum
dependent parts of the form factors.  It should be noted that in our
manipulations we retained both the symmetric and the antisymmetric
parts of $d_{\mu\nu}$.  However, as is expected the results only
depend on the symmetric part of $d_{\mu\nu}$.  Now the total
contribution on the EDM form factor can be found by adding
(\ref{vf230}) and (\ref{vf330}) as
\begin{eqnarray}
\Gamma_{2\,p\cdot d\cdot p}+\Gamma_{3\,p\cdot d\cdot
p}&=&\frac{2e^2}{(4\pi^2)}\int
dxdydz\delta(x+y+z-1)\frac{2m}{((1-z)^2m^2)^2}\bar{u}(p')\{\nonumber\\&
+&(x-y)z^2y(1-y)-z^3y^2-z^3(x+y)(z-2y+2)\nonumber\\&-&
z^2xy(z-2y+2)+2y[3z(1-y)(z+y-1) +(1-y)^2]\nonumber\\&-&2xz^3\}p\cdot
d^s\cdot p(p+p')^{\mu}\ga_5u(p),\label{vft}
\end{eqnarray}
which after performing the integrals on the Feynman parameters,
leads to
\begin{eqnarray}
\Gamma_{2\,p\cdot d\cdot p}+\Gamma_{3\,p\cdot d\cdot
p}&=&-\frac{2e^2}{(4\pi^2)}\bar{u}(p')(\frac{275}{18m^3})p\cdot
d^s\cdot p(p+p')^{\mu}\ga_5u(p)+IR,\label{vft}
\end{eqnarray}
in which  $IR$ stands for the infrared terms.  By comparing
(\ref{vft}) and (\ref{edm}), one can easily see that
\begin{eqnarray}
F_4&=&-\frac{275\al}{18\pi}\frac{p.d^s.p}{m^2}. \label{f4}
\end{eqnarray}


\end{document}